\documentclass[12pt,preprint]{aastex}
\slugcomment{Draft}

\shorttitle{Standing Accretion Shocks}
\shortauthors{Laming}

\begin{document}

\title{Analytic Approach to the Stability of Standing Accretion Shocks:
Application to Core-Collapse Supernovae}


\author{J. Martin Laming\altaffilmark{1}}


\altaffiltext{1}{E. O Hulburt Center for Space Research, Naval
Research Laboratory, Code 7674L, Washington DC 20375
\email{laming@nrl.navy.mil}}

\begin{abstract}
We explore an analytic model of the accretion shock in the post bounce phase
of a core-collapse supernova explosion. We find growing oscillations of the shock
in the $l=1$ and $l=2$ modes, in agreement with a variety of existing numerical simulations.
For modest values of the ratio of the outer accretion shock to that of the inner
boundary to the shocked flow, the instability appears to derive from the growth
of trapped sound waves, whereas at higher values, postshock advection clearly plays
a role. Thus the model described here may relate to the different mechanisms of instability
recently advocated by Blondin \& Mezzacappa, and by Foglizzo and collaborators.
\end{abstract}


\keywords{supernovae: general---hydrodynamics---instabilities---shock waves}

\section{Introduction}
There is a growing consensus that a large fraction of core-collapse supernovae,
possibly including the whole subset of Type Ib/c events, undergo significantly
asymmetrical explosions. While clearly rotation and magnetic fields may play
some role in this, instability associated with the stalled shock that forms at a radius
of 100-300 km and lasts for a few hundred microseconds post bounce offers the
intriguing possibility of generating asymmetrical explosions from a symmetrical
progenitor, since such instability is usually dominated by low mode $l=1$, 2
oscillations of the shock front. The original speculation for the origin of such
oscillations \citep{herant92,herant95} in terms of convection behind the shock,
following \citet{chandrasekhar61} has been revised in recent years.
\citet{foglizzo05} demonstrate
that post shock advection can stabilize convective instability (\citet{chandrasekhar61}
only deals with a static case), and that convection
alone cannot produce the dominant $l=1$,2 modes.
In adiabatic simulations of gas with
polytropic index 4/3, which as shown by \citet{janka01} is a good description of shocked
inner regions of a core-collapse supernova, \citet{blondin03} were the
first to associate the instability with the shock itself. \citet{blondin05} attributed this to the
growth of trapped sound waves, travelling essentially
laterally around the central accretor. This must be a purely shock driven instability,
since convection is absent, and hence qualitatively different to the original speculations
\citep{herant92,herant95}. On the other hand, \citet{foglizzo06}
and \citet{galletti05} attribute this shock-driven instability to an ``advective-acoustic'' cycle,
originally introduced by \citet{foglizzo02} in the context of black hole accretion flows.
Vorticity perturbations generated at the distorted shock front advect inwards, and couple with
outgoing sound waves at the inner boundary of the postshock accretion flow. Upon reaching the outer
shock, these sound waves further distort the shock front, leading to a positive feedback.
While the mechanisms of \citet{blondin05} and \citet{foglizzo02} appear
to produce similar outcomes in non-rotating cases, indeed \citet{foglizzo06} actually confirm the
numerical values of frequencies and growth rates measured by \citet{blondin05}, it is possible
the presence of rotation might lead to substantial differences, which we discuss further below.

Instabilities of the accretion shock also are observed in simulations with a more accurate treatment
of supernova microphysics, with a realistic equation of state, neutrino transport
and attendant nuclear reactions, and more physically motivated boundary conditions.
\citet{scheck04} and \citet{scheck06} investigate the role
of such instabilities in generating pulsar natal kicks, while \citet{kifonidis05}
study the ejection of metal clumps to the outer regions of supernova ejecta.
\citet{burrows05} also see the standing accretion shock instability 100 - 300 ms
postbounce. The important feature of their simulation, though, is a core gravity wave
at late times that transfers acoustic power to the shock to continue powering the
explosion, the initial perturbation for which is possibly seeded by the accretion
shock instability. \citet{burrows05} also favor the interpretation of \citet{foglizzo06} of
an advective-acoustic cycle, though in this and the other more realistic simulations cited above,
the exact mechanism of instability is difficult to identify. To avoid this problem,
\citet{ohnishi05} performed numerical simulations based on unperturbed spherically
symmetrical shock accretion flows given by \citet{yamasaki05}, a sort of ``middle ground''
between the adiabatic flows of \citet{blondin03} and \citet{blondin05} and the full
supernova simulations of other workers. \citet{ohnishi05} clearly identify and
measure growth rates for $l=1$ and 2 mode instabilities. The real parts of the
frequencies scale with $\int _{r_i}^{r_s}\left(1/c_s +1/v_r\right) dr$, the integral between the
inner radius of the accretion flow and the shock of the sum of the inverses of the
sound speed and advection speed, suggesting that an advective-acoustic mechanism,
as advanced by \citet{foglizzo02} and \citet{foglizzo06}, is at work.
However \citet{ohnishi05} only perturb the
radial velocity component of the initially spherical symmetrical unperturbed flow, which
may unduly influence their results, compared to allowing perturbations in polar and azimuthal
directions as well. \citet{yamasaki06} include rather more physics, and again based on the frequencies
at which modes grow, conclude that an advective-acoustic cycle is at work.

In this work we present another approach to the problem. Following
the work of \citet{vishniac89} on planar shocks in the interstellar
medium, we derive an approximate dispersion relation for
oscillations of the accretion shock by adding the effect of a
gravitational field and spherical geometry. In the case that the
postshock advection is neglected, a quartic equation (actually a
quadratic in $\omega ^2$ where $\omega $ is the frequency of
oscillation) results, becoming a quintic equation in $\omega$ with
the inclusion of advection. We find that in both cases for
$\gamma\sim 4/3$, for all except the highest values of the ratio of
the shock radius to the radius of the inner boundary, modes with
$l=1$ grow the fastest (in fact for cases where the postshock radial
velocity, $u_r=0$, no other nonradial modes grow). This suggests
that for $l=1$ in $\gamma =4/3$ gas in this regime, postshock
advection is not crucial to the operation of the instability. Thus
in general these results will support the conclusions of
\citet{blondin05} that the instability proceeds by the growth of
trapped waves in this regime. Nevertheless, regions of $l$ and
$\gamma$ parameter space exist where we find modes that are only
unstable in the presence of postshock advection, which we interpret
as an advective-acoustic cycle, as suggested by \citet{galletti05}
and \citet{foglizzo06}. In both cases, the frequencies of the
growing modes are similar, suggesting that frequency alone is not a
good discriminator of the mechanism of instability. As will be seen
below, this might be expected in a comparison of a perturbation
advected radially between the shock and inner boundary, and a sound
wave traveling essentially laterally around the shock.

\section{Analytic Theory}
We consider an unperturbed model in which spherically accreting plasma is decelerated at a
spherically symmetrical shock, before accreting onto a protoneutron star. The accretion
shock is at radius $r_s$, and we take an inner boundary at $r_i$, maintained at constant
pressure, where plasma cools and
decouples from the postshock flow. The postshock flow is modeled by the Bernoulli equation,
as given in Appendix A. Gas with polytropic index $\gamma < 1.5$ decelerates, and gas
with $\gamma > 1.5$ accelerates away from the shock towards the inner boundary. We shall
be solely concerned with $\gamma < 1.5$. The dominant variations with radius are in pressure
and density, as shown in Figure 1.

To treat the perturbation,
we follow in large part the methods and notation of \citet{vishniac89} who derive an approximate
dispersion relation for application to shocks with arbitrary postshock structure.
In spherical coordinates, we write the hydrodynamic equations;
an equation of continuity and momentum
equations in the $r$, $\theta$ and $\phi$ directions as

Continuity:
\begin{eqnarray}
{\partial\left(\rho +\Delta\rho\right)\over\partial t}+\left(u_r+v_r\right)
{\partial\left(\rho +\Delta\rho\right)\over\partial r}+
{v_{\theta}\over r}{\partial\left(\rho +\Delta\rho\right)
\over\partial\theta}+
{\left(u_{\phi}+v_{\phi}\right)\over r\sin\theta}
{\partial\left(\rho +\Delta\rho\right)\over\partial\phi}&\cr
+\left(\rho +\Delta\rho
\right)\left[{1\over r\sin\theta}{\partial\sin\theta v_{\theta}\over\partial\theta}
+{1\over r\sin\theta}{\partial\left(u_{\phi}+v_{\phi}\right)\over
\partial\phi }+{1\over r^2}{\partial\left(r^2\left(u_r+v_r\right)\right)\over
\partial r}\right]=0&
\end{eqnarray}

Radial Force:
\begin{eqnarray}
\left(\rho +\Delta\rho\right)\left[{\partial\left(u_r+v_r\right)\over\partial t}
+\left(u_r+v_r\right){\partial\left(u_r+v_r\right)\over\partial r} +
\left(u_{\phi}+v_{\phi}\over r\sin\theta\right){\partial\left(u_r+v_r\right)
\over\partial\phi}-{\left(u_{\phi}+v_{\phi}\right)^2\over r}\right]&\cr =
-{\partial\left(P+\Delta P\right)\over\partial r}+{\partial\over\partial r}
\left[GM\left(\rho +\Delta\rho\right)\over r\right]&
\end{eqnarray}

Poloidal Force:
\begin{eqnarray}
\left(\rho +\Delta\rho\right)\left[{\partial v_{\theta}\over\partial t}
+u_r{\partial v_{\theta}\over\partial r}+{v_{\theta}u_r\over r}-\cot\theta
{u_{\phi}^2\over r}--2\cot\theta {u_{\phi}v_{\phi}\over r} +{u_{\phi}\over
r\sin\theta }{\partial v_{\theta}\over\partial\phi }\right]=&\cr
-{1\over r}{\partial\left(P+\Delta P\right)\over\partial\theta }&
\end{eqnarray}

Azimuthal Force:
\begin{eqnarray}
\left(\rho +\Delta\rho\right)\left[{\partial\left(u_{\phi}+v_{\phi}\right)
\over\partial t}
+\left(u_r+v_r\right){\partial\left(u_{\phi}+v_{\phi}\right)\over\partial r} +
{v_{\theta}\over r}{\partial u_{\phi}\over\partial\theta }+\cot\theta {u_{\phi}
v_{\theta}\over r}\right]&\cr+
\left(\rho +\Delta\rho\right)\left[
\left(u_{\phi}+v_{\phi}\over r\sin\theta\right){\partial\left(u_{\phi}+v_{\phi}
\right)
\over\partial\phi}-{\left(u_{\phi}+v_{\phi}\right)\left(u_r+v_r\right)
\over r}\right]=
-{1\over r\sin\theta }{\partial\left(P+\Delta P\right)\over\partial\phi},&
\end{eqnarray}
where $\rho$ and $\Delta\rho$ are the initial and perturbed density respectively,
$u_r$ and $v_r$ are the initial and perturbed radial velocities and $u_{\phi}$ and
$v_{\phi}$ and the initial and perturbed azimuthal velocities. The initial
poloidal velocity is zero, and its perturbed value is $v_{\theta}$. $P$ and $\Delta P$
are initial and perturbed pressures, and $GM$ is the gravitational constant times
the enclosed mass which is dominated by that of the protoneutron star.
Writing $\Delta\rho /\rho =\delta$, and assuming
$\delta\propto\exp\left(i\omega t+\int\lambda dr\right)Y_{lm}\left(\theta ,\phi\right)$ where
$\lambda =\lambda\left(r\right)$ we derive a linearized
continuity equation,
\begin{equation}
i\omega ^{\prime}\delta+v_r\left({1\over L}+{2\over r}\right)+u_r\lambda\delta
+{imv_{\phi}\over r\sin\theta}+{\partial v_r\over\partial r}
+{1\over r\sin\theta}{\partial\sin\theta v_{\theta}\over\partial\theta}=0,
\end{equation}
where $\omega ^{\prime}=\omega +mu_{\phi}/r\sin\theta =\omega +m\Omega$ where
$\Omega = u_{\phi}/r\sin\theta $ and $L=\left(1-\gamma\right)r$ is the density scale length,
defined by $\partial\ln\rho /\partial r = 1/L$. The $\phi$ derivatives of terms in $u_{\phi}$
and $v_{\phi}$ have been taken to yield the $m$ dependence, but the $\theta$ derivative of the
term in $v_{\theta}$ remains explicit. Below (in section 3.2) we will assume
$u_{\phi}\propto r\sin\theta$ rendering $\Omega$ constant.

The radial force equation is linearized as follows. The zero order equation
gives
\begin{equation}
\rho {\partial u_r\over\partial t}+u_r{\partial u_r\over\partial r}
+{u_{\phi}\over r\sin\theta}{\partial u_r\over\partial\phi }=-{\partial P\over
\partial r}+{\partial\over\partial r}\left(GM\rho\over r\right).
\end{equation}
If $\partial u_r/\partial\phi =0$ and $\partial u_r/\partial t=0$, and
$\gamma =1$ ($u_r=0$) or $\gamma =1.5$ ($\partial u_r/\partial r=0$), then
equation 6 evaluates to zero. We will assume that this is approximately true for
all $1\le\gamma\le 1.5$. (Certainly $u_r\partial u_r/\partial r < 0$ throughout this
range, whereas for $\gamma > 1.5$, $u_r\partial u_r/\partial r > 0$).
This helps ensure that for all cases where the postshock gas
decelerates, the radial $l=0$ mode will be stable, in agreement with the conclusions
of \citet{nakayama92}, and \citet{burrows93} and \citet{yamasaki05}, where we are assuming a neutrino
luminosity below the critical value in these last two references.
Taking $\Delta P=P\delta$ (i.e. an approximately
isothermal perturbation; this gives a considerable simplification, see Appendix B)
the equation to first order in small quantities becomes
\begin{eqnarray}
&{\partial\over\partial r}\left[u_rv_r\exp\int i{\omega ^{\prime}\over u_r}dr\right]
=\cr
&\left[{2u_{\phi}v_{\phi}\over r}-{v_{\theta}\over r}{\partial u_r\over\partial
\theta}
+\delta\left[\left({GM\over r}-c_s^2\right)\lambda
+{2u_r^2+u_{\phi}^2\over r}+{u_r^2\over L}
\right]\right]\exp\int i{\omega ^{\prime}\over u_r}dr .\cr
\end{eqnarray}
We note that
\begin{equation}\nonumber
{d\over dr}\left[{1\over i\omega ^{\prime}/u_r +\lambda}\exp\int\lambda+i\omega ^{\prime}/u_rdr
\right]=\left[{-d\lambda /dr+i\left(\omega /u_r^2\right) \partial u_r/\partial r\over
\left(i\omega ^{\prime}/u_r+\lambda\right)^2}+1\right]
\exp\int\lambda+i\omega ^{\prime}/u_rdr.
\end{equation}
For a global mode we expect $\lambda$ to be of order $1/r$, (in fact
for $\gamma =4/3$, $\lambda\simeq 2/r$; see equations 16 below). If
$\omega$ represents laterally propagating sound waves, $\omega$ is
in the range $c_s/r - c_s/\left|L\right|$, whereas $\omega\sim
2\pi\left(r/u_r+r/c_s\right)$ for radially propagating vorticity
perturbations. In either case, with $\partial u_r/\partial r=\left(u_r/r\right)\left(3-2\gamma\right)/\left(
\gamma -1\right)$ from Appendix A, ${-d\lambda /dr+i\left(\omega /u_rr\right)\left(3-2\gamma\right)/\left(
\gamma -1\right)\over\left(i\omega
^{\prime}/u_r+\lambda\right)^2} <<1$ and we may put
\begin{equation}\nonumber
{d\over dr}\left[{1\over i\omega ^{\prime}/u_r +\lambda}\exp\int\lambda+i\omega ^{\prime}/u_rdr
\right]\simeq\exp\int\lambda+i\omega ^{\prime}/u_rdr.
\end{equation}
In equation 7 we neglect terms in $u_r^2/c_s^2$, $u_{\phi}^2/c_s^2$ (of order $\left(\gamma -1\right)/2\gamma$)
and integrate using equation 9, further assuming
$\left|i\omega ^{\prime}/u_r+\lambda\right| >>\left|1/r\right|$, to get
\begin{equation}
v_r={2u_{\phi}v_{\phi}\over i\omega ^{\prime}r}-{v_{\theta}\over
i\omega ^{\prime}r}{\partial u_r\over\partial\theta}+{\delta\over
i\omega ^{\prime}+u_r\lambda}\lambda\left({GM\over r}-c_s^2\right).
\end{equation}
With $u_{\phi}=0$ and $\partial u_r/\partial\theta =0$, equation 7 gives
\begin{equation}
{\partial v_r\over\partial r}u_r=\delta\left({GM\over r}-c_s^2\right)\lambda -i\omega v_r -{\partial u_r
\over\partial r}v_r=\left(u_r\lambda -{\partial u_r\over\partial r}\right)v_r,
\end{equation}
where we have substituted from equation 10 for $v_r$ to simplify the right hand side. Taking
$u_r\propto r^{\left(3-2\gamma\right)/\left(\gamma -1\right)}$ from Appendix A, we derive
\begin{equation}
{\partial v_r\over\partial r}=\left(\lambda - {3-2\gamma\over\gamma -1}{1\over r}\right)v_r.
\end{equation}
Expressions for $v_{\theta}$ and
$v_{\phi}$ are easily derived in the limits $u_{\phi}\rightarrow
0$ or $u_r\rightarrow 0$.
Staying with the nonrotating case for the time being, we give for $u_{\phi}=0$:
\begin{eqnarray}
v_{\theta}&=-{c_s^2\over r}{\partial\delta\over\partial\theta}{1\over
i\omega +\lambda u_r}\cr
v_{\phi}&=-i{mc_s^2\over r\sin\theta }{\delta\over i\omega +\lambda u_r}.
\end{eqnarray}
where the primes have been dropped from $\omega$.

Substituting our expression for $v_r$, $\partial v_r/\partial r$, $v_{\theta}$ and $v_{\phi}$ into the
linearized continuity equation (5) we find
\begin{equation}
i\omega\delta +{\delta\lambda\over i\omega +\lambda u_r}
\left({GM\over r}-c_s^2\right)
\left({1\over L}+{4\gamma -5\over\gamma -1}{1\over r}+\lambda\right)+u_r\lambda\delta
+{c_s^2\over i\omega +\lambda u_r}{l\left(l+1\right)\delta\over r^2}=0,
\end{equation}
where the $m$ dependence disappears (as it should in spherical symmetry) using properties of the
spherical harmonics.
This can be solved for $\lambda$ to give
\begin{eqnarray}
&\lambda _{\pm}=-\left({1\over 2L}+{4\gamma -5\over\gamma -1}{1\over 2r}\right)-{iu_r\omega\over GM/r-c_s^2}\cr
&\pm\left({1\over 2L}+{4\gamma -5\over\gamma -1}{1\over 2r}\right)\sqrt{1+4{iu_r\omega\over\left(GM/r-
c_s^2\right)\left(1/L+\left(4\gamma -5\right)/\left(\gamma -1\right)/r\right)}+4{\omega ^2-l\left(l+1\right)c_s^2/r^2
\over\left(1/L+\left(4\gamma -5\right)/\left(\gamma -1\right)/r\right)^2\left(GM/r-c_s^2\right)}}.
\end{eqnarray}
We write (neglecting terms in $u_r^2/c_s^2$)
\begin{eqnarray}
\lambda _+\lambda _-&={l\left(l+1\right)c_s^2/r^2 -\omega ^2\over GM/r
-c_s^2}\cr
\lambda_++\lambda _-&=-\left({1\over L}+{4\gamma -5\over\gamma -1}{1\over r}\right)-{2iu_r\omega\over
GM/r-c_s^2}.
\end{eqnarray}

As discussed by \citet{vishniac89}, a third solution exists,
with $\delta = 0$ everywhere. In that case $v_r\propto\exp\left(-i\int\omega /u_r dr\right)/u_r$,
and $v_{\theta}$ and $v_{\phi}\propto\exp\left(-i\int\omega /u_r dr\right)/u_rr$. The linearized
continuity equation (5) then gives
\begin{equation}
v_r\left({1\over L} +{2\over r}\right)+{\partial v_r\over\partial r}
+\vec{\nabla _{\perp}}\cdot\vec{v_{\perp}}=0.
\end{equation}
Taking $\partial v_r/\partial r=-iv_r\omega /u_r >> v_r\left(1/L+2/r\right)$ we find
\begin{equation}
v_r={u_r\over i\omega}\vec{\nabla _{\perp}}\cdot\vec{v_{\perp}}.
\end{equation}

The derivation of the dispersion relation from the boundary conditions in terms of these three solutions
is carried out in Appendix B, and given by equation B6. We use $\lambda _{\pm}^{\prime}=
\lambda _{\pm}/\left(1+\lambda _{\pm}u/i\omega\right)$, with $\lambda _s$ and $\lambda _i$ evaluated
at the outer shock and inner boundary respectively in terms of $u_{rs}$ and $u_{ri}$, the advection
velocities at these locations (see Appendix B for more details). Then in equation B6 we approximate
\begin{eqnarray}
&{\lambda _{+i}^{\prime}-\lambda _{-i}^{\prime}\beta ^Q\over 1-\beta ^Q}=
-\left({1\over 2L}+{4\gamma -5\over\gamma -1}{1\over 2r}\right)\left(1-Q{1+\beta ^Q\over 1-\beta ^Q}\right)
+{iu_{ri}\over 2\omega}\left({1\over 2L}+{4\gamma -5\over\gamma -1}{1\over 2r}\right)^2
\left(1+Q{1+\beta ^Q\over 1-\beta ^Q}\right)\cr
&+{iu_{ri}\omega\over GM/r-c_s^2}{1\over Q}{1+\beta ^Q\over
1-\beta ^Q} -{iu_{ri}\over\omega}{l\left(l+1\right)c_s^2/r^2\over GM/r-c_s^2}\cr
&{\lambda _{is}^{\prime}-\lambda _{+s}^{\prime}\beta ^Q\over 1-\beta ^Q}=
-\left({1\over 2L}+{4\gamma -5\over\gamma -1}{1\over 2r}\right)\left(1+Q{1+\beta ^Q\over 1-\beta ^Q}\right)
+{iu_{rs}\over 2\omega}\left({1\over 2L}+{4\gamma -5\over\gamma -1}{1\over 2r}\right)^2
\left(1-Q{1+\beta ^Q\over 1-\beta ^Q}\right)\cr
&-{iu_{rs}\omega\over GM/r-c_s^2}{1\over Q}{1+\beta ^Q\over
1-\beta ^Q} -{iu_{rs}\over\omega}{l\left(l+1\right)c_s^2/r^2\over GM/r-c_s^2}
\end{eqnarray}
and
\begin{eqnarray}
&{\lambda _+\lambda _-^{\prime}-\lambda _-\lambda_+^{\prime}\beta ^Q\over 1-\beta ^Q}=
{l\left(l+1\right)c_s^2/r^2-\omega ^2\over GM/r-c_s^2}\left[1-\left({1\over L}+
{4\gamma -5\over\gamma -1}{1\over r}\right)\left({iu_{rs}\over 2\omega}\left(1+Q{1+\beta ^Q\over 1-\beta ^Q}\right)
-{iu_{ri}\over 2\omega}\left(1-Q{1+\beta ^Q\over 1-\beta ^Q}\right)\right)\right]\cr
&+\left({1\over L}+{4\gamma -5\over\gamma -1}{1\over r}\right){i\omega\over 2\left(GM/r-c_s^2\right)}
\left[u_{rs}\left(2-Q{1+\beta ^Q\over 1-\beta ^Q}+{1\over Q}{1+\beta ^Q\over 1-\beta ^Q}\right)
u_{ri}\left(2+Q{1+\beta ^Q\over 1-\beta ^Q}-{1\over Q}{1-\beta ^Q\over 1-\beta ^Q}\right)\right]
\end{eqnarray}
where we have put
\begin{eqnarray}Q^{\prime}&=\sqrt{1+4i{u_r\omega\over\left(GM/r-c_s^2\right)
\left(1/L+\left(4\gamma -5\right)/\left(\gamma -1\right)/r\right)}+
4{\omega ^2-l\left(l+1\right)c_s^2/r^2
\over\left(1/L+\left(4\gamma -5\right)/\left(\gamma -1\right)/r\right)^2\left(GM/r-c_s^2\right)}}\cr
&\simeq Q  +{2i\over Q}{u\omega\over
\left(GM/r-c_s^2\right)/\left(1/L+\left(4\gamma -5\right)/\left(\gamma -1\right)/r\right)}
\end{eqnarray}and neglect terms of $u^2/c_s^2$ and higher.
These then give
\begin{eqnarray}
&\omega ^4+\omega ^3\left[iu_{rs}\left({1\over L}+{5\over 2r}\right)-i{u_{ri}\over aL}\right]
{1\over Q}{1+\beta ^Q\over 1-\beta ^Q}\cr
&-\omega ^2\left[{l\left(l+1\right)\over r^2}v_su_r
+{GM/r-c_s^2\over aL}\left({1\over L}+{5\over 2r}\right)\right]\cr
&+\omega ^2\left[{GM/r-c_s^2\over 2}\left({1\over L}+{4\gamma -5\over\gamma -1}{1\over r}\right)
\left({1\over L}+{5\over 2r}\right)\left(1+Q{1+\beta ^Q\over 1-\beta ^Q}\right)+
{GM/r-c_s^2\over 2}\left({1\over L}+{4\gamma -5\over\gamma -1}{1\over r}\right)
{1\over aL}\left(1-Q{1+\beta ^Q\over 1-\beta ^Q}\right)\right]\cr
&+\omega\left[iu_{rs}\left({1\over L}+{5\over 2r}\right)l\left(l+1\right){c_s^2\over r^2}
+i{u_{ri}\over aL}l\left(l+1\right){c_s^2\over r^2}+
i{u_{ri}\over aLQ}{1+\beta ^Q\over 1-\beta ^Q}
l\left(l+1\right){v_su_{rs}\over r^2}\right]\cr
&+\omega\left[-{i\over 2}\left({GM\over r}-c_s^2\right)\left({1\over L}+{4\gamma -5\over
\gamma -1}{1\over r}\right)^2\left(u_{rs}\left({1\over L}+{5\over 2r}\right)\left(1-Q{1+\beta ^Q\over
1-\beta ^Q}\right)+{u_{ri}\over L}\left(1+Q{1+\beta ^Q\over 1-\beta ^Q}\right)\right)\right]\cr
&+\omega\left[{{GM\over r}-c_s^2\over 2aL}\left({1\over L}+{4\gamma -5\over
\gamma -1}{1\over r}\right)\left({1\over L}+{5\over 2r}\right)\left(3iu_{rs}+{iu_{rs}\over Q}{1+\beta ^Q\over
1-\beta ^Q}+iu_{ri}+{iu_{ri}\over Q}{1+\beta ^Q\over
1-\beta ^Q}\right)\right]\cr
&+{GM/r-c_s^2\over aL}\left({1\over L}+{5\over 2r}\right)
{l\left(l+1\right)c_s^2\over r^2}
-{GM/r-c_s^2\over 2aL}\left({1\over L}+{4\gamma -5\over\gamma -1}{1\over r}\right)
\left({1\over L}+{5\over 2r}\right)
{l\left(l+1\right)v_su_{rs}\over r^2}
\left(1-Q{1+\beta ^Q\over 1-\beta ^Q}\right)\cr
&+{1\over\omega}\left[{GM/r-c_s^2\over 2al}\left({1\over L}+{5\over 2r}\right)l\left(l+1\right){c_s^2\over r^2}
\left({1\over L}+{4\gamma -5\over\gamma -1}{1\over r}\right)\left(-iu_{rs}\left(1+Q{1+\beta ^Q\over 1-\beta ^Q}
\right)-iu_{ri}\left(1-Q{1+\beta ^Q\over 1-\beta ^Q}\right)\right)\right]\cr
&{1\over\omega}\left[-{l\left(l+1\right)v_su_{rs}\over r^2}
\left(i{u_{ri}\over 2aL}\left(GM/r-c_s^2\right)\left({1\over L}+{4\gamma -5\over\gamma -1}{1\over r}\right)^2
\left(1+Q{1+\beta ^Q\over 1-\beta ^Q}\right)+i{u_{ri}\over aL}{l\left(l+1\right)c_s^2\over r^2}\right)\right]=0
\end{eqnarray}

If $u_r=0$, equation 22 is very similar to equation 13a of
\citet{vishniac89}, and is identical if we take
$r\rightarrow\infty$, $L\rightarrow -L$ (to agree with their sign
convention), $l\left(l+1\right)/r^2\rightarrow k^2$
and $\gamma\rightarrow 1$ (i.e. $v_su_r\rightarrow c_s^2$). In their case the limit $Q\rightarrow 1$
gives a dispersion relation similar to the exact solution in the thin shock limit \citep{vishniac83},
and identical to this limit if terms describing the evolution of the shock are neglected from the
exact solution. They then proceed to argue that taking $Q\ne 1$ gives a suitable approximate
dispersion relation away from the thin shock limit. Our case is particularly suitable since the
standing accretion shock can be treated as an equilibrium state \citep{burrows93}, i.e. it does
not evolve in our approximation.

Just taking $u_r=0$, equation 20 yields
a quadratic equation in $\omega ^2$ for which simple solutions
exist. \citet{vishniac89} find growing shock oscillations in the case of decelerating
shocks, with a post shock rarefaction. In our case, the growing oscillations exist
for density increasing with distance behind the shock. This difference arises
because the square of the sound speed, $c_s^2\rightarrow c_s^2 -GM/r$ in
our case, and $GM/r > c_s^2$. When $u_r=0$ (except in terms in $v_su_r$) and taking $Q\rightarrow
1$,
\begin{eqnarray}
&\omega ^2={l\left(l+1\right)v_su_r\over 2r^2}-\left({GM\over r}
-c_s^2\right)\left({1\over 2L}+{5\over 4r}\right){4\gamma -5\over\gamma -1}{1\over r}\cr &\pm
\left\{{l\left(l+1\right)v_su_r\over 2r^2}-\left({GM\over r}
-c_s^2\right)\left({1\over 2L}+{5\over 4r}\right){4\gamma -5\over\gamma -1}{1\over r}\right\}\sqrt{1-W}.
\end{eqnarray}
where
\begin{equation}W=4{\left({GM\over
r}-c_s^2\right){1\over L}\left({1\over L}+{5\over 2r}\right){l\left(l+1\right)v_su_r\over r^2}
\over \left\{{l\left(l+1\right)v_su_r\over r^2}-\left({GM\over r}
-c_s^2\right)\left({1\over L}+{5\over 2r}\right){4\gamma -5\over\gamma -1}{1\over r}\right\}^2}
\end{equation}
When $1>>W$,
this gives a pair of sound waves for given $l$ with
\begin{equation}\omega ^2\simeq l\left(l+1\right)v_su_r/r^2-
\left(GM/r^2-c_s^2/r\right)\left(1/L +5/2r\right){4\gamma -5\over\gamma -1}
\end{equation}
taking the positive sign, and the negative sign gives a pair of
gravity waves with
\begin{equation}\omega ^2\simeq 2{\left({GM\over
r}-c_s^2\right){1\over L}\left({1\over L}+{5\over 2r}\right){l\left(l+1\right)v_su_r\over r^2}
\over \left\{{l\left(l+1\right)v_su_r\over r^2}-\left({GM\over r}
-c_s^2\right)\left({1\over L}+{5\over 2r}\right){4\gamma -5\over\gamma -1}{1\over r}\right\}},
\end{equation}
where $a$ is a constant coming into our definition of the interior boundary condition, as explained in
Appendix B. For $1 < W$, we have four roots of
the form $\omega = \pm \left(f+ig\right)$ and $\omega = \pm\left(f-ig\right)$, and all roots
have a combined character of acoustic and gravity waves. Reinstating the terms in odd powers
of $\omega$ with nonzero $u_r$, the roots separate into pairs of what may be termed acoustic and
gravity waves, although the difference in frequency between them is generally rather small and both pairs
are of presumably mixed character,
and one imaginary (i.e. purely growing or damping) mode. Assuming a value of $Q$, we solve the quintic equation
numerically. Results for each growing mode are then iterated until the value of $Q$ produced by the
frequency and wavenumber matches the input. Results of this procedure for growing modes are given in
Table 1 for $\gamma =4/3$, $a=1.5$, and in Table 2 for $\gamma =1.36$, $a=2$, the values of $a$ being chosen
to reproduce as far as possible the stability properties of the radial mode.
We compare in each case the results of the full quintic equation with a simpler
case taking $u_{rs}=u_{ri}=0$.

With $\gamma =4/3$, taking $GM/rc_s^2\simeq 1/\left(\gamma -1\right)=3$ (from equation A2 neglecting
$u_r^2$), and without advection, we find growth for $l=1$ for all values
of $r_s/r_i\le 0.1$. Growth at $l=1$ at $r_s/r_i\rightarrow\infty$ and at $l=2$ for all
$r_s/r_i$ require non-zero advection terms in the dispersion relation, though the $l=2$ mode is
stable as $r_s/r_i\rightarrow\infty$, even in the presence of advection. The $l=0$ mode is
marginally stable in the absence of advection. With advection included, it is stable for the
lowest $r_s/r_i$, but otherwise unstable. Taken together, there is reasonable qualitative
agreement with \citet{blondin05}, but quantitatively, the model is less stable than it should be.
Growth rates are too high, and the $l=0$ mode becomes unstable too quickly as $r_s/r_i$ increases.
Better qualitative and quantitative agreement is found in the model with $\gamma =1.36$ and $a=2$ given
in Table 2., and also
illustrated in Figure 2. The stability of $l=0$ is improved, and advection becomes relatively more
necessary to the instability of $l=1$. These results have a simple interpretation in terms of the
mechanisms put forward by \citet{blondin05} and \citet{foglizzo06}. Where advection is not necessary
to the instability, i.e. for the $l=1$ mode at low values of $r_s/r_i$, the wave growth must be
due to trapped sound waves similarly to the simulations of \citet{blondin05}. The inclusion of
advection in these cases generally increases the growth rate, either due to a separate contribution
of an ``advective-acoustic'' nature or because of the effect of postshock advection on the propagation
of sound waves. In cases where no instability occurs unless advection terms are present, the growth
must be due to an advective-acoustic cycle similar to \citet{foglizzo06}. In our model, these cases
include all the $l=2$ modes, as well as the $l=1$ modes for large values of $r_s/r_i$.

These results and interpretation add a new twist to the conclusions of \citet{foglizzo06}. These
authors demonstrate that for $r_s/r_i > 10$ the advective-acoustic cycle is the cause of the
$l=1$ instability. Their analysis makes use of a WKB approximation that is not valid for lower
$r_s/r_i$. However they proceed to argue that since growth rates and frequencies vary smoothly as
one goes to lower values of $r_s/r_i$, the mechanism of instability should remain the same. However
our results suggests that even though the eigenvalues vary smoothly, the mechanism of instability changes
to more closely resemble that suggested by \citet{blondin05}.

\section{Discussion}
\subsection{Preamble}
A number of simplifying
assumptions have been introduced into the analysis to make the algebra tractable. We have
assumed an isothermal perturbation, used a greatly simplified interior boundary condition, and
have dropped all terms in the advection velocity of order higher than $u_r/c_s$.
Nevertheless our model appears to capture the basic features of the numerical simulations of
\citet{blondin05}, {\em and} the analytic theory of \citet{foglizzo06} and \citet{galletti05},
in the appropriate regions of parameter space. Looking at the values of $\lambda _{\pm}$ given
by equation 15, we find when $\omega ^2\sim l\left(l+1\right)c_s^2/r^2$,
$\left|\lambda _-\right|>>\left|\lambda _+\right|$, $\lambda _->0$
and $\lambda _+{>\atop <}0$ for $Q{<\atop >}1$. Consequently, $\lambda _-$ gives a density perturbation
$\delta =\delta\rho /\rho$ that is greatest at the outer shock and considerably smaller at
the inner boundary, while $\lambda _+$ gives a density perturbation that varies much less with
radius, but can be largest at the outer shock or at the inner boundary. When $l=0$, $\lambda _+ <0$ and
$\left|\lambda _-\right|\sim\left|\lambda _+\right|$. In this case the radial mode can have a large
maximum of $\delta$ at the inner boundary. Given the ad hoc nature of our inner boundary condition (see
Appendix B), we might expect less accurate results for the radial mode than for nonradial modes, which
does indeed seem to be the case. The vorticity (calculated from equations 13)
increases inwards approximately as $\delta /r^2$.
The third solution with $\delta =0$ everywhere must represent
a purely vortical perturbation.

\subsection{The Mechanism of Instability}
So far we have developed a dispersion relation from the equations of hydrodynamics,
and found instabilities through the existence of complex roots. We have not made any
comment on the precise mechanism of the instability, other than to comment that sound/gravity
waves appear to grow. The vortical-acoustic
cycle has been explained by \citet{foglizzo02}. A distortion of the accretion
shock produces a vorticity perturbation that is advected inwards. Upon reaching the inner
boundary, the vorticity perturbation produces an upward moving sound wave, which reinforces
the shock distortion when it arrives there, producing a net increase in the original
perturbation. Over many cycles, exponential growth ensues.

Sound waves produced at the outer shock, in a realistic model, will not propagate towards
the center due to the effects of refraction. Instead, they will propagate around the
circumference of the shock. \citet{blondin05} speculate that sound waves produced at one
position on the shock surface by a density inhomogeneity will propagate around until they meet
on the opposite side, where their excess pressure produces a shock distortion that sends
another pair of sound waves back again. It is not immediately clear without detailed
calculation that such a process
should produce a growing (as opposed to damping) oscillation.
We speculate that growth must also be aided by the flow of plasma
in through the accretion shock. In Figure 3 we show flowlines for shocked plasma
in an accretion shock with a 10\% $l=1$
modulation of the shock radius. When the shock is shifted up by the instability, incoming
plasma in the equatorial regions is diverted downwards by the now oblique shock. This non-radial
flow of shocked plasma enhances the oscillation and leads to further wave growth. This aspect
is similar to the thin shell overstability of decelerating shocks
\citep{vishniac83,vishniac89}, (although \citet{velikovich05} put forward a different view),
though in these cases the pressure produced by gravitational confinement plays no role, and
overstability exists for a postshock density gradient oppositely directed to our case. Thus
both the advective-acoustic (or vortical-acoustic) instability and the growth due to trapped
sound waves have their origins in perturbations of either vorticity or pressure in
the post shock flow produced by the distorted shock front.

\subsection{The Effect of Rotation}
There is considerable interest in the effects of rotation on core-collapse explosions. In
particular, in which direction with respect to the rotation axis should one expect the strongest growth?
This would have obvious implications for the directions of pulsar kicks, assuming the such instabilities
are the correct mechanism. In the case of an advective-acoustic
instability, we should expect the decrease in the postshock advection in the plane of
rotation to give stronger instability and hence strongest growth along the axis of rotation.

In the case of instability driven by trapped sound waves we can sketch out
the effect of including a nonzero $u_{\phi}$ in the forgoing
analysis. We derive the extra terms to be included in the dispersion relation in the limits
$u_r\rightarrow 0$ and $u_{\phi}^2/c_s^2\rightarrow 0$.
We also neglect any distortion of the shock front by the rotation. Such considerations require more detailed
physics, neutrino cooling near the inner boundary at a minimum \citep{blondin05}, and ideally
also neutrino luminosities and mass accretion rates \citep[see e.g.][]{burrows93,yamasaki05},
and are beyond the scope of this paper. This keeps $u_{\theta}=0$, which should
be true in any case if $u_r=0$, but even in the absence of advection, a distortion of the shock
front may change the boundary condition leading to equation B2. The perturbed velocities are
\begin{eqnarray}
v_r&={\delta\lambda\over i\omega ^{\prime}}\left({GM\over r}-c_s^2
\right)-{2m\Omega c_s^2\delta\over ir\omega ^{\prime 2}}\cr
v_{\theta}&=-{2mc_s^2\Omega\delta\cot\theta\over ir\omega ^{\prime 2}}
-{c_s^2\over ir\omega }{\partial\delta\over\partial\theta}\cr
v_{\phi}&=-{mc_s^2\delta\over\omega ^{\prime}r\sin\theta}+
{2\Omega\delta\lambda\sin\theta\over\omega ^{\prime 2}}\left({GM\over r}-
c_s^2\right) -{2\Omega c_s^2\cos\theta\over r\omega ^{\prime 2}}{\partial\delta\over\partial\theta}.
\end{eqnarray}
With
\begin{equation}
{\partial v_r\over\partial r}=-{v_r\over r}+{\delta\left(GM/r-c_s^2\right)\lambda ^2\over i\omega ^{\prime}},
\end{equation}
the linearized continuity equation becomes
\begin{eqnarray}
&-\omega ^{\prime 2}+\left\{\left({GM\over r}-c_s^2\right)\lambda-
{2m\Omega c_s^2\over r\omega ^{\prime}}\right\}
\left({1\over L}+{1\over r}\right)+\lambda ^2\left({GM\over r}-c_s^2\right)
+{l\left(l+1\right)c_s^2\over r^2}+{2mc_s^2\Omega\over r^2\omega ^{\prime}}
-{2mc_s^2\Omega\lambda\over r\omega ^{\prime}}=0,
\end{eqnarray}
after dividing through by $\delta$, with solutions for $\lambda _{\pm}$
\begin{eqnarray}
&\lambda _++\lambda _-=-\left({1\over L}+{1\over r}\right)+{2\Omega m\over
r\omega ^{\prime}}\cr
&\lambda _+\lambda _-={l\left(l+1\right)c_s^2/r^2 -\omega ^{\prime 2}
\over GM/r-c_s^2}-{2m\Omega c_s^2\over rL\omega ^{\prime}}
{1\over GM/r-c_s^2}.
\end{eqnarray}
We again have $v_r\propto\exp\int -i\omega ^{\prime}/u_rdr$ if $\delta =0$, so equation (14)
still holds. Then, with unchanged boundary conditions,
the dispersion relation is the same as before (with $u_r=0$ and $\omega\rightarrow
\omega ^{\prime}$ in equation B6), but with the extra
terms (neglecting the change $1/L +\left(4\gamma -5\right)/\left(\gamma -1\right)/r\rightarrow 1/L+1/r$,
which in any case is trivial for $\gamma =4/3$)
\begin{eqnarray}
&\omega ^{\prime}{m\Omega\over r}\left({GM\over r}-c_s^2\right)\left({2\over L}+{5\over 2r}Q{1+
\beta ^Q\over 1-\beta ^Q}\right)\cr &
-{l\left(l+1\right)v_su_r\over r^2}{m\Omega\over\omega ^{\prime} r}\left({GM\over rL}-{c_s^2\over L}\right)
\left(1-Q{1+\beta ^Q\over 1-\beta ^Q}\right)
-\left({GM\over rL}-{c_s^2\over L}\right)\left({1\over L}+{5\over 2r}\right){2m\Omega c_s^2\over rL
\omega ^{\prime}}.
\end{eqnarray}
The new quintic equation is again solved numerically, with results given in Table 3 for the
$m=1$ component of the $l=1$ mode. We generally find higher
growth rates increasing with rotation rate for the $m=1$ ($m=0$ remains unchanged),
which would indicate higher growth in the plane of rotation.
This happens even in cases
where in the absence of rotation, the advective-acoustic instability seems to be dominant.
We speculate that rotation increases the growth rate of trapped sound waves for nonzero $m$,
so much so that they can grow even in cases with zero growth in the absence of rotation.

At higher rotation rates, the distortion of the shock front cannot be neglected. We expect the
equilibrium shock position to be oblate, i.e. bulging out at the equator as demonstrated quantitatively by
\citet{yamasaki05}. We emphasize again that our results are not applicable in this regime.

\section{Conclusions}
Using heuristic arguments based on \citet{vishniac89}, and a few other approximations, we have
derived an approximate dispersion relations for oscillations of a standing spherical accretion shock.
The pursuit of an analytic treatment
further through the problem than has previously been attempted arguably allows greater physical insight,
albeit at the expense of reduced quantitative accuracy.
While our model might not be considered much more than a ``toy model'', it does capture most of
the features of previous works of the instability of such shocks.  It offers a
potential resolution to the mild disagreement between the works of \citet{blondin05} and
\citet{foglizzo06}. We find results consistent with both works, but in different regions of parameter
space. For low values of $r_s/r_i$, the growth of trapped waves dominates in the $l=1$,
and appears to be the most important mechanism. The inclusion of post shock advection gives a small
change, usually an increase, in the growth rate. At higher $l$, or $r_s/r_i$, postshock
advection appears to become crucial to the instability, which leads us to interpret it as an
advective-acoustic cycle. In a treatment of $\gamma =4/3$ gas in
core collapse supernovae, \citet{galletti05} only find advective-acoustic instability for
$r_s/r_i > 3.5$, with the growth rate increasing out to $r_s/r_i \sim 10$. \citet{foglizzo06} are only able
to demonstrate the operation of the advective-acoustic cycle for $r_s/r_i \ge 10$, due to the nature of
the approximations involved, but conjecture that
it should also operate at lower $r_s/r_i$. Our models agree with their demonstrations at large
shock radii, but suggest that at smaller shock radius, the instability should change to resemble that
observed by \citet{blondin05}.

The precise mechanism of instability probably becomes more significant in the rotating case.
The advective-acoustic instability must grow preferentially along the rotation axis, where the
postshock advection is not affected by the rotation. By contrast, we conjecture in section 3.3 that
rotation should enhance the growth of trapped sound/gravity waves principally in the plane of
rotation, leading to stronger growth in directions perpendicular to the rotation axis. Such effects
may have important consequences for pulsar natal kicks, if these derive from hydrodynamic kick
mechanisms.

\acknowledgements
This work has been supported by the Chandra GI Program, the NASA LTSA and APRA programs, and by basic
research funds of the Office of Naval Research. I acknowledge a scientific discussion
with Ethan Vishniac, and an extremely careful and constructive reading of the paper by an anonymous
referee.

\appendix
\section{Bernoulli Equation for the Postshock Flow}
Here we summarize the model for the postshock flow in spherical accretion used
by \citet{blondin03}. The flow is given by the Bernoulli equation
\begin{equation}
u_r^2 +{2\gamma\over\gamma -1}{P\over\rho }-{2GM\over r}=0
\end{equation}
and just behind the shock itself the jump conditions give the velocity, density and
pressure as
\begin{eqnarray}
&u_r=-{\gamma -1\over\gamma +1}\sqrt{2GM\over r_s}\cr
&\rho={\gamma +1\over\gamma -1}{\dot{M}\over 4\pi}\sqrt{r_s\over 2GM}
{1\over r_s^2}\cr
&P={2\over\gamma +1}{\dot{M}\over 4\pi}\sqrt{2GM\over r_s}{1\over r_s^2}.
\end{eqnarray}
Evaluating $P/\rho ^{\gamma} = {\rm constant}$ from the jump conditions and
substituting gives
\begin{equation}
r^{\prime}u_r^{\prime 2}+{4\gamma\over\left(\gamma +1\right)
\left(\gamma -1\right)}\left(\gamma -1\over\gamma +1\right)^{\gamma}
r^{\prime 3-2\gamma}u_r^{\prime 1-\gamma}-1=0
\end{equation}
where $r^{\prime}=r/r_s$ and $u^{\prime}=u_r\sqrt{r_s/2GM}$. Neglecting the first
term (i.e the kinetic energy), $u_r\propto r^{\left(3-2\gamma\right)/\left(
\gamma -1\right)}$ and hence $\rho\propto r^{-1/\left(\gamma -1\right)}$. These
results are exact for $\gamma =5/3$. In this approximation, the density decreases
monatonically with increasing radial distance, and $L$ defined above is $L=r
\left(1-\gamma\right)$. We note that for
$\gamma\rightarrow 1$, close to the shock, the density may increase with increasing
radial distance, i.e. a postshock rarefaction exists. For $\gamma =1$, this region
extends between $0.75r_s < r < r_s$, and disappears for $\gamma > 1.1$.

\section{Boundary Conditions and Dispersion Relation}
We recapitulate and modify very slightly the approach of \citet{vishniac89}.
Let $B_{\pm}$ be the amplitudes of the $\lambda _{\pm}$ oscillations of the accretion shock.
Then at the accretion shock, $\delta\left(0\right)=-\delta r_s/L-5\delta r_s/2r_s -2\delta v_s/v_s\simeq
-\delta r_s/L-5\delta r_s/2r_s=-\left(1/L+5/2r\right)v_r/i\omega$,
which comes from the radial variation of the shock ram pressure. We follow \citet{vishniac89} and
neglect the term in $\delta v_s$, which renders the perturbation isothermal.
\citet{blondin05} also comment that
the term in $\delta v_s$ is smaller by a factor of a few than the term in $\delta r_s$ in determining
the shock ram pressure. Hence
\begin{equation}
B_++B_-=\left({\left(GM/r-c_s^2\right)\over\omega ^2}\left(B_+\lambda _{+s}^{\prime}
+B_-\lambda _{-s}^{\prime}\right)+{u_r\nabla _{\perp}\cdot\vec{v_{\perp}}\over\omega ^2}\right)\left(
{1\over L}+{5\over 2r}\right),
\end{equation}
where $\lambda _{\pm}^{\prime}=\lambda _{\pm}/\left(1+\lambda _{\pm}u/i\omega\right)$, and the subscript
$s$ indicates that this is evaluated at the outer shock.

Also at the accretion shock, $v_{\perp}\left(r_s\right)=v_s\vec{\nabla _{\perp}}\delta r_s=
\left(v_s/i\omega\right)\vec{\nabla _{\perp}}v_r\left(r_s\right)$, which after some
rearrangement gives
\begin{equation}
\vec{\nabla _{\perp}}\cdot\vec{v_{\perp}}=-{v_s\left(GM/r-c_s^2\right)\over\omega ^2}\nabla _{\perp}^2
\left(B_+\lambda _{+s}^{\prime}+B_{-s}\lambda _-^{\prime}\right)
-{v_su_r\over\omega ^2}\nabla _{\perp}^2\left(\vec{\nabla _{\perp}}\cdot
\vec{v_{\perp}}\right).
\end{equation}
Eliminating $\vec{\nabla _{\perp}}\cdot\vec{v_{\perp}}$ between equations B1 and B2 gives
\begin{eqnarray}
&B_+\left(\omega ^2L^{\prime}\left(1-{v_su_rl\left(l+1\right)\over\omega ^2r^2}\right)-
\left({GM\over r}-c_s^2\right)\lambda _{+s}^{\prime}\right)+\cr
&B_-\left(\omega ^2L^{\prime}\left(1-{v_su_rl\left(l+1\right)\over\omega ^2r^2}\right)-
\left({GM\over r}-c_s^2\right)\lambda _{-s}^{\prime}\right)=0
\end{eqnarray}
where $\nabla _{\perp}^2B_{\pm} = -l\left(l+1\right)B_{\pm}/r^2$,
$\nabla _{\perp}^2\left(\vec{\nabla _{\perp}}\cdot
\vec{v_{\perp}}\right) = -l\left(l+1\right)\left(\vec{\nabla _{\perp}}\cdot
\vec{v_{\perp}}\right)/r^2$ and $1/L^{\prime}=1/L+5/2r$.

The interior boundary, and hence the boundary conditions here, are less well defined
in our model. We are assuming that the region where strong neutrino cooling sets in and causes
the accreting matter to decouple from the hot postshock flow will give a boundary from which
waves can reflect. \citet{vishniac89}, considering a plane parallel shock, assume a constant pressure
boundary condition, $\delta =-\delta r_i/L=-v_r/i\omega L$. We assume something similar;
\begin{equation}\delta\left(r_i\right)=B_+\exp\int _{r_i}^{r_s}\lambda _+dr
+B_-\exp\int _{r_i}^{r_s}\lambda _-dr
=-{v_r\left(r_i\right)\over i\omega aL}=-{\delta\left(r_i\right)
\left(GM/r-c_s^2\right)\over\omega ^2aL}.
\end{equation}
We do not take the local value of $L$ in
this boundary condition, but take the same value as assumed above at the forward shock, modified by a
constant $a$ which we vary to find the best match to the $l=0$ stability. In any case,
$\delta\left(r_i\right)=-v_r\left(r_i\right)/i\omega aL\left(r_s\right)
<<-v_r\left(r_i\right)/i\omega L\left(r_i\right)$ as it
must be, because the background density increases significantly over that predicted by the Bernoulli model
due to radiative cooling by neutrino emission. We prefer to cast the interior boundary condition in terms
of $L$ and $r$ evaluated at the outer shock, because of the algebraic simplification it produces, and
the somewhat easier conditions required to keep the $l=0$ mode
stable. It is clear from the work of various authors \citet[e.g.][]{nakayama92,burrows93,yamasaki05}
that this must be true. These last two references, in particular, include considerably more physics,
and are able to predict the shock radius, rather than just specifying it as we do, but they
only treat radial stability. Consequently we have
\begin{equation}
B_+\exp\int _{r_i}^{r_s}\left(\lambda _+ -\lambda _-\right)dr
\left\{1-{1\over \omega ^2aL}\left[\left({GM\over r}-c_s^2\right)\lambda _{+i}^{\prime}\right]\right\}
+B_-\left\{1-{1\over \omega ^2aL}\left[\left({GM\over r}-c_s^2\right)\lambda _{-i}^{\prime}\right]\right\}=0,
\end{equation}
where the $\lambda _{\pm i}$ are evaluated with $u$ at the inner boundary, but otherwise with $L$ and $r$
at the outer shock.

Now eliminating $B_+/B_-$ between
equations B3 and B5 gives the dispersion relation:

\begin{eqnarray}
&\omega ^4\left[1-\beta ^Q\right]-\omega ^2{l\left(l+1\right)\over r^2}
v_su_r\left[1-\beta ^Q\right]
-\omega ^2\left({GM\over r}-c_s^2\right){1\over aL}\left[\lambda _{+i}^{\prime}
-\lambda _{-i}^{\prime}\beta ^Q\right]\cr
&-\omega ^2\left({GM\over r}-c_s^2\right)\left({1\over L}+{5\over 2r}\right)
\left[\lambda ^{\prime}_{-s}-\lambda ^{\prime}_{+s}\beta ^Q\right]
+{l\left(l+1\right)\over r^2}v_su_r\left({GM\over r}-c_s^2\right)
{1\over aL}\left[\lambda _{+i}^{\prime}-\lambda _{-i}^{\prime}\beta ^Q\right]\cr
&+\left({GM\over r}-c_s^2\right)^2
{1\over aL}\left({1\over L}+{5\over 2r}\right)
\left[\lambda _{+i}^{\prime}\lambda _{-s}^{\prime}-\lambda _{+s}^{\prime}\lambda _{-i}^{\prime}\beta ^Q\right]
=0
\end{eqnarray}
where
\begin{equation}
\beta ^Q=\exp\int _{r_i}^{r_s}-\left(\lambda _+-\lambda _-\right)dr
=\exp\int _{r_i}^{r_s}\left({1\over L}+{4\gamma -5\over\gamma -1}{1\over r}\right)Qdr
\simeq\left(r_i\over r_s\right)^{{6-4\gamma\over\gamma -1}Q}.
\end{equation}
For $1<\gamma <1.5$, $\beta\rightarrow 0$ as $r_i/r_s\rightarrow 0$, and as $\gamma\rightarrow 1$
for fixed $r_s/r_i$.

\clearpage
\begin{table}[t]
\begin{center}
\caption{Eigenvalues of growing modes $0\le l\le 2$: $\gamma =4/3$, $a=1.5$}
\begin{tabular}{lrrrrrr}
\tableline\tableline
 & \multicolumn{3}{c}{no advection}&\multicolumn{3}{c}{advection}\\
$r_s/r_i$  & $l=0$ & $l=1$ & $l=2$ & $l=0$ & $l=1$ & $l=2$   \\
\tableline
2  & $\pm 0.569$ & $\pm 0.525$  &  & $\pm 0.569$ & $\pm 0.546$& $\pm 0.771$\\
   &  & $-0.242i$&  & $+0.034i$ & $-0.284i$& $-0.326i$\\
 \\
3  & $\pm 0.391$ & $\pm 0.443$ &  & $\pm 0.447$ & $\pm 0.520$ &$\pm 0.730$\\
  &  &$-0.253i$ &   & $-0.090i$& $-0.305i$ &$-0.312i$\\
 \\
5  & $\pm 0.257$ & $\pm 0.387$& & $\pm 0.453$ & $\pm 0.511$ & $\pm 0.707$\\
  & & $-0.248i$ &   & $-0.155i$ & $-0.310i$ & $-0.297i$\\
 \\
10  & $\pm 0.145$ &  $\pm 0.342$& & $\pm 0.479$  & $\pm 0.516$ & $\pm 0.693$\\
 &  & $-0.238i$ &  &  $-0.178i$ & $-0.311i$ &$-0.287i$\\
 \\
$\infty$ & &&&$\pm 0.511$ & $\pm 0.532$& \\
 & &&& $-0.199i$ &  $-0.311i$ & \\
\tableline
\end{tabular}
\tablecomments{Real and imaginary frequencies are in units of $\left|c_s/L\right|$. Growing modes
have a negative imaginary part.}
\end{center}
\end{table}

\clearpage
\begin{table}[t]
\begin{center}
\caption{Eigenvalues of growing modes $0\le l\le 2$: $\gamma =1.36$, $a=2$}
\begin{tabular}{lrrrrrr}
\tableline\tableline
 & \multicolumn{3}{c}{no advection}&\multicolumn{3}{c}{advection}\\
$r_s/r_i$  & $l=0$ & $l=1$ & $l=2$ & $l=0$ & $l=1$ & $l=2$   \\
\tableline
2  & $\pm 0.506$ & $\pm 0.495$  &  & $\pm 0.523$ & $\pm 0.490$& $\pm 0.829$\\
   &  & $-0.193i$&  & $+0.010i$ & $-0.222i$& $-0.173i$\\
 \\
3  & $\pm 0.347$ & $\pm 0.404$ &  & $\pm 0.353$ & $\pm 0.454$ &$\pm 0.818$\\
  &  &$-0.175i$ &   & $+0.500i$& $-0.228i$ &$-0.171i$\\
 \\
5  & $\pm 0.219$ & & & $\pm 0.264$ & $\pm 0.425$ & $\pm 0.811$\\
  & &  &   & $-0.058i$ & $-0.220i$ & $-0.176i$\\
 \\
10  & $\pm 0.044$ &  & & $\pm 0.287$  & $\pm 0.401$ & $\pm 0.806$\\
 &  &  &  &  $-0.11i$ & $-0.201i$ &$-0.194i$\\
 \\
$\infty$ & &&&$\pm 0.343$ & $\pm 0.383$& \\
 & &&& $-0.147i$ &  $-0.119i$ & \\
\tableline
\end{tabular}
\tablecomments{Real and imaginary frequencies are in units of $\left|c_s/L\right|$. Growing modes
have a negative imaginary part.}
\end{center}
\end{table}

\clearpage
\begin{table}[t]
\begin{center}
\caption{Eigenvalues of $l=1$, $m=1$ $\gamma =1.36$, $a=2$}
\begin{tabular}{lrrrrrr}
\tableline\tableline
 & \multicolumn{6}{c}{$u_{\phi}$}\\
$r_s/r_i$  & 0.0& 0.05& 0.1& 0.2& 0.3& 0.4\\
\tableline
2  & -0.490 & -0.498  & -0.502 & -0.511 & -0.520& -0.527\\
 & $-0.222i$& $-0.231i$ & $-0.242i$& $-0.264i$& $-0.282i$ & $-0.298i$\\
 \\
   &  0.490 &  0.489 &  0.484& 0.473& 0.461& \\
 & $-0.222i$& $-0.205i$ & $-0.190i$& $-0.154i$& $-0.091i$ &\\
 \\
 \\

10  & 0.401 & 0.430 & 0.433& 0.441  & 0.449 & 0.456\\
 & $-0.201i$ & $-0.224i$ & $-0.241i$ & $-0.264i$& $-0.281i$ & $-0.294i$ \\
 \\
    & -0.401& -0.425& -0.430&  -0.464& -0.483& -0.495\\
 & $-0.201i$ & $-0.171i$ & $-0.115i$ & $-0.045i$ & $-0.032i$& $-0.027i$ \\

\tableline
\end{tabular}
\tablecomments{Real ($\omega ^{\prime}=\omega +m\Omega$)
and imaginary frequencies are in units of $\left|c_s/L\right|$. Growing modes
have a negative imaginary part. For $m=-1$, real parts take opposite signs.}
\end{center}
\end{table}

\clearpage
\begin{figure}
\epsscale{0.6} \plotone{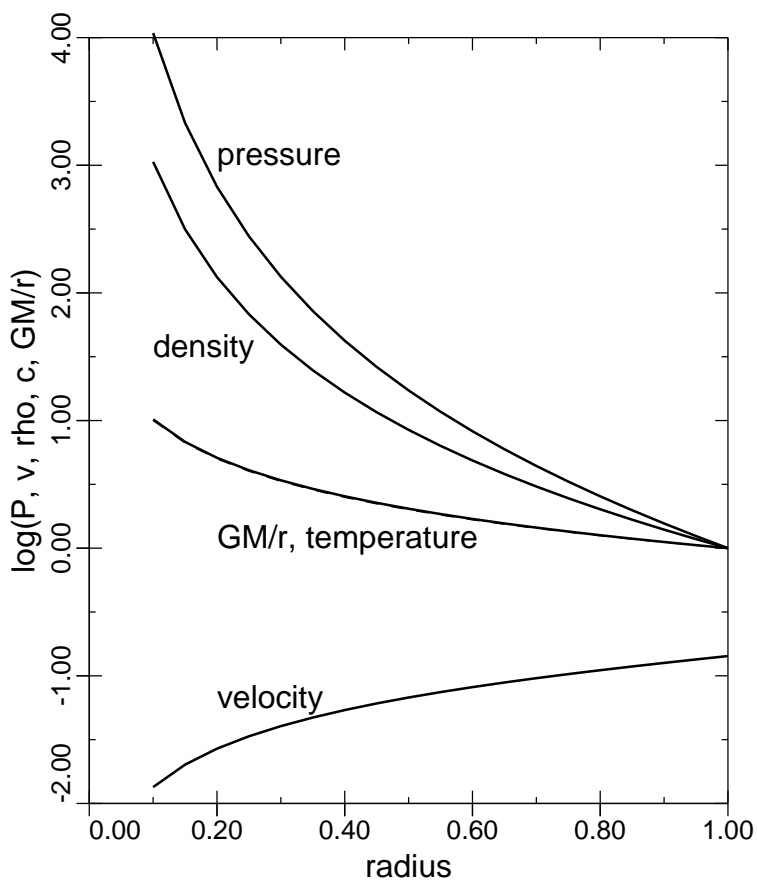} \figcaption[f1.eps]{Variation behind
the accretion shock of the pressure, density, inward flow velocity,
sound speed and gravitational potential for a $\gamma =4/3$ model.
All quantities are normalized to unity at the accretion shock and
radius unity, except the inward flow velocity, which is 1/7, the
shock velocity being unity at radius =1. The pressure and density
exhibit much bigger variations with distance than any other
quantity, motivating us to simplify the problem by taking the sound
speed, inward flow velocity and gravitational potential to be
constant throughout the shocked plasma. \label{fig1}}
\end{figure}

\begin{figure}
\epsscale{0.6} \plotone{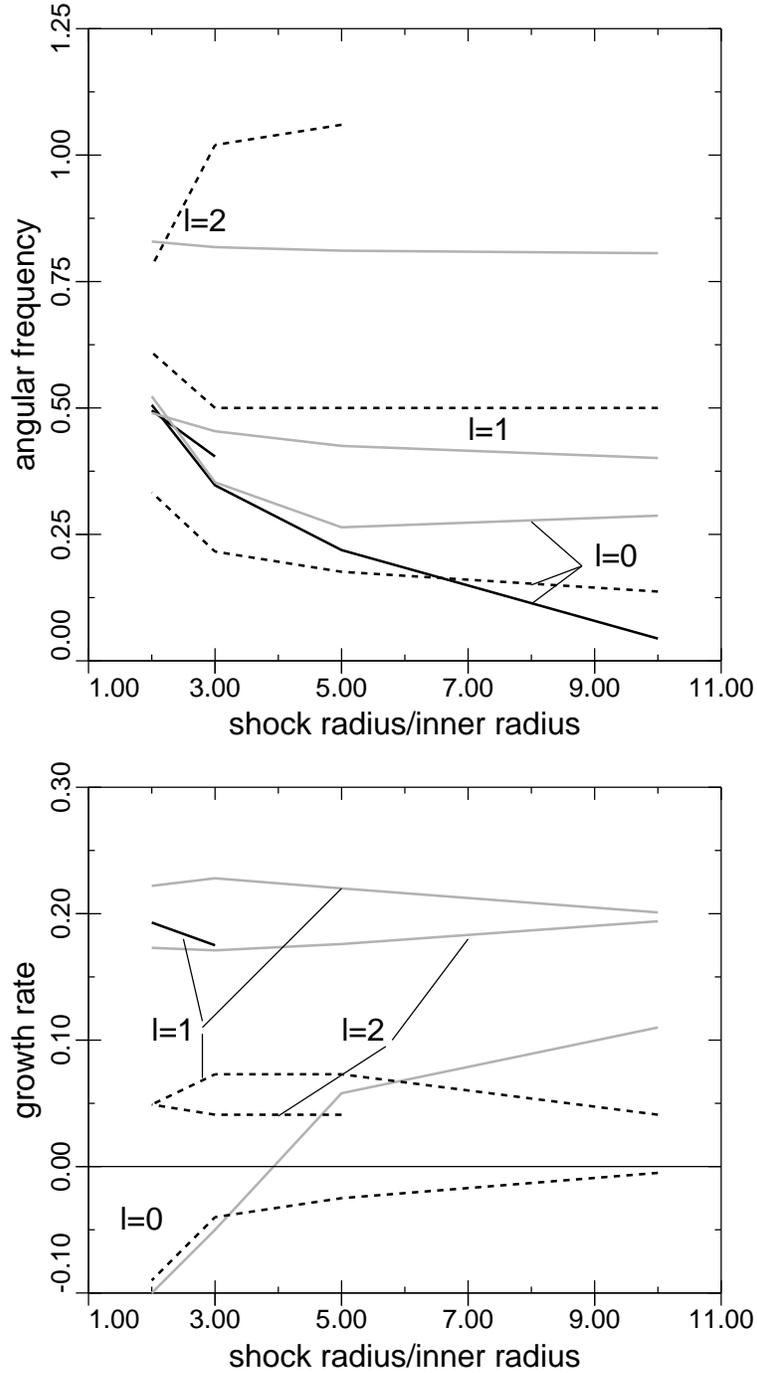} \figcaption[f2.eps]{Angular
frequencies and growth rates of unstable modes plotted against
$r_s/r_i$. Solid lines give present results, (black without
advection; grey with advection), and the dashed lines are those from
\citet{blondin05}, all in units of $c_s/L$ evaluated close to the
accretion shock. The growth rate of the $l=0$ mode without advection
is 0; it is only marginally stable, and is not plotted.
\label{fig2}}
\end{figure}

\begin{figure}
\epsscale{0.6} \plotone{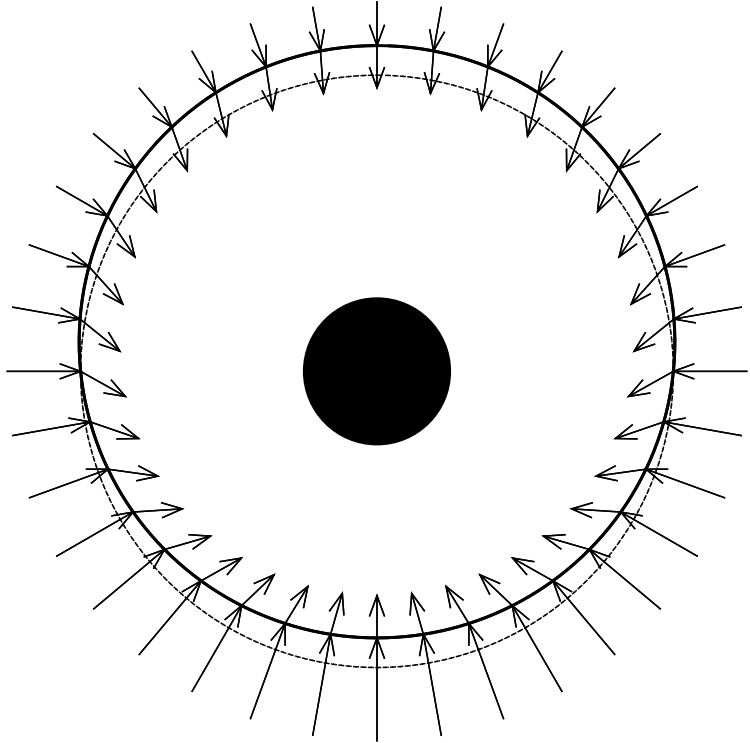} \figcaption[f3.eps]{Post shock
advection velocity flow lines for an accretion shock with 10\% $l=1$
modulation of radius for $\gamma =4/3$ (upper panel). The shock
perturbed upwards (solid line) from its equilibrium position (dashed
line) has its surface distorted such that incoming plasma flows
preferentially towards the lower density and pressure region at the
bottom of the figure. This helps induce the growing oscillation.
\label{fig3}}
\end{figure}

\end{document}